# Vacuum Thermal Switch Made of Phase Transition Materials Considering Thin Film and Substrate Effects

Yue Yang,  Soumyadipta Basu, and Liping Wang[*]

*School for Engineering of Matter, Transport, and Energy*
*Arizona State University, Tempe, AZ 85287*
Corresponding author: liping.wang@asu.edu

**ABSTRACT**

In the present study, we demonstrate a vacuum thermal switch based on near-field thermal radiation between phase transition materials, i.e., vanadium dioxide ($VO_2$), whose phase changes from insulator to metal at 341 K. Similar modulation effect has already been demonstrated and it will be extended to thin-film structure with substrate in this paper. Strong coupling of surface phonon polaritons between two insulating $VO_2$ plates significantly enhances the near-field heat flux, which on the other hand is greatly reduced when the $VO_2$ emitter becomes metallic, resulting strong thermal switching effect. Fluctuational electrodynamics predicts more than 80% heat transfer reduction at sub-30-nm vacuum gaps and 50% at vacuum gap of 1 micron. By replacing the bulk $VO_2$ receiver with a thin film of several tens of nanometers, the switching effect can be further improved over a broad range of vacuum gaps from 10 nm to 1 μm. In addition, for the purpose of more practical setup in experiments and applications, the $SiO_2$ substrate effect is also considered for the structure with thin-film emitter or receiver.

**Nomenclature**

$c$        speed of light in vacuum, $2.997 \times 10^{8}$ m/s

$d$        vacuum gap distance, m

$k_{\mathrm{B}}$       Boltzmann constant, $1.38 \times 10^{-23}$ m$^2$ kg s$^{-2}$ K$^{-1}$

$q''$       radiative heat flux, W/m$^2$

$R$        reflection coefficient of a thin film

$r$        Fresnel reflection coefficient for a single interface

$s$        exchange function, m$^{-1}$

$T$        temperature, K

$t$        thin film thickness, m

*Greek symbols*

$\beta$        parallel component of wavevector, m$^{-1}$

$\gamma$        vertical component of wavevector, m$^{-1}$

$\varepsilon$        dielectric function

$\Theta$        Planck oscillator, J

$\hbar$        reduced Planck constant, $1.055 \times 10^{-34}$ J s

$\Phi$        thermal switching factor

$\omega$        angular frequency, rad/s

*Subscripts*

0, 1, 2, 3, 4        layer 0, 1, 2, 3, 4

evan      evanescent waves

prop      propagating waves

on        "on" mode

off       "off" mode

E        extraordinary component of dielectric function

H        high-temperature medium

$i$        medium $i$

L        low-temperature medium

O        ordinary component of dielectric function

*Superscripts*

$E$        extraordinary component of dielectric function

O        ordinary component of dielectric function

p        p polarization

s        s polarization

## 1. Introduction

Thermal switch, like electric switch having "on" and "off" modes, plays a crucial role in controlling heat flow, especially for microelectronics applications [1-3]. Cho et al. [4, 5] developed a thermal switch based on arrays of liquid–metal micro-droplets to control heat transfer. DNA amplification via polymerase chain reaction (PCR) [6, 7] is another important application. In addition, thermal switch was used to control the heat into or out of a micro heat engine designed by Whalen et al [8]. However, the thermal switches referred above were all based on making and breaking thermal contact between two interfaces to realize the thermal switching effect, while the repeating mechanical movement limits the lifetime of these switches.

The researches on nanoscale thermal radiation have showed that the radiant energy exchanges in near field can exceed that predicted by Planck's law by a few orders of magnitude [9-13]. The surface plasmon or phonon polaritons (SPhPs) effect [14, 15] can significantly enhance the near-field thermal radiation. These facts make it possible to use thermal radiation for thermal management by manipulating heat flow in the near-field. Recently, by taking advantage of the characteristics of near-field thermal radiation, Zhu et al. [16] described a bistable thermal switch made of two SiC plates with different polytypes placed in close proximity with a nanometer vacuum gap, while high temperature above 800 K is required to achieve thermal switch effect. Besides, the modulation of near-field radiative heat transfer between phase transition $VO_2$ has been demonstrated. van Zwol et al. [17] calculated the near-field radiative heat transfer coefficients between two insulating $VO_2$ surfaces or two metallic $VO_2$ surfaces with a vacuum gap distance much smaller than dominant thermal wavelength. The former can be more than one order of magnitude higher than the latter at extremely small gap distance $d < 10$ nm, due to the strong SPhP coupling between two insulating $VO_2$-vacuum interfaces that significantly enhances the heat transfer. They [18] also experimentally demonstrated the modulation of near-field radiative heat transfer between a glass microsphere and a phase transition $VO_2$ film with an AFM-tip based technique. Recently, radiation-based vacuum thermal rectifiers and transistors have been also been designed by employing phase change of $VO_2$ [19-21] or AIST [22] to achieve radiative thermal modulation.

In the present study, we will first quantitatively show the near-field thermal switching effect between two semi-infinite $VO_2$ plates, one of which is maintained at room temperature as an insulator while the other serves as an emitter to experience insulator-metal transition when temperature increases. The penetration depth of near-field thermal radiation inside the insulating $VO_2$, which is a uniaxial medium, will be then discussed. After that, we will extend the study to a 4-layer vacuum thermal switch, by considering the receiving $VO_2$ as a free-standing thin film, in order to understand the effect of film thickness on the near-field thermal switching. Besides a vacuum substrate, a real film substrate made of $SiO_2$ is also investigated to study the practical performance of the proposed near-field thermal switch with the consideration of both emitting and receiving $VO_2$ film thicknesses.

## 2. Theoretical Background

### 2.1 Optical Properties of Phase Transition $VO_2$

$VO_2$ is a kind of insulator-metal transition material at the temperature of 341 K. Crystalline $VO_2$ with (200) orientation [23, 24] is considered in this study, which is a uniaxial insulator with its optical axis vertical to the surface when temperature is below 341 K, and an isotropic metal when temperature otherwise. The dielectric properties of crystalline $VO_2$ below or above phase transition temperature were experimentally studied in Ref. [23]. When VO₂ is a uniaxial insulator, the dielectric function can be expressed as a tensor

$$\bar{\bar{\varepsilon}} = \begin{pmatrix} \varepsilon_O & 0 & 0 \\ 0 & \varepsilon_O & 0 \\ 0 & 0 & \varepsilon_E \end{pmatrix} \tag{1}$$

where the Lorentz model $\varepsilon(\omega) = \varepsilon_\infty + \sum_{j=1}^{N} \dfrac{S_j \omega_j^2}{\omega_j^2 - i\gamma_j\omega - \omega^2}$ is used to describe the dielectric function. Seven phonon modes and one band-structure mode are used to fit the component of dielectric function in the plane vertical to the optical axis, i.e., ordinary dielectric function $\varepsilon_O$, while eight phonon modes and one band-structure mode are used to fit the component of dielectric function parallel to the optical axis, i.e., extraordinary dielectric function $\varepsilon_E$. The value for each parameter, such as phonon strength $S_j$, phonon frequency $\omega_j$, and damping coefficient $\gamma_j$, can be found in Ref. [23]. The real and imaginary parts of the dielectric function for both ordinary and extraordinary components of insulating VO₂ are shown in Fig. 1. It can be clearly seen that there exist multiple phonon modes for insulating VO₂ in the infrared region, which could possibly excite SPhP for enhancing near-field radiative heat transfer.

On the other hand, when VO₂ is an isotropic metal, the Drude model $\varepsilon_m = -\omega_p^2 \varepsilon_\infty / \left( \omega^2 - i\omega\omega_c \right)$ is used to describe the dielectric function, where $\omega$ is the angular frequency, $\varepsilon_\infty = 9$ is the highi-freuqency constant, $\omega_p = 8000$ cm⁻¹, and $\omega_c = 10000$ cm⁻¹ are plasma and collision frequency, respectively [23]. As shown in Fig. 1, the real part of dielectric function for metallic VO₂ little depends on the frequency, while the imaginary part decreases monotonically when the angular frequency increases from 2×10¹³ rad/s to 2×10¹⁴ rad/s.

Qazilbash et al. [24] observed that the VO₂ transition from insulator to metal is not completed instantly once the temperature is above 341 K, but via a gradual process, in which the insulator and metal phases co-exist over a temperature range from 341 K to 350 K. The VO$_2$ during the transition process can be treated as a mixture of insulating and metallic VO$_2$, whose filling fractions as well as the resulting effective dielectric function of the mixture changes with temperature. To facilitate the discussion in the present study, we will simply assume that VO$_2$ completely changes to the metallic phase when the temperature is 1 K above 341 K. The dielectric functions of insulating and metallic VO$_2$ are also treated to be independent on the temperature.

## 2.2 *Proposed Vacuum Thermal Switch and Fluctuational Electrodynamics*

Figure 2 depicts the schematic of a vacuum thermal switch made of five layers including the emitter substrate (denoted as layer 0), an emitting VO$_2$ film (layer 1), vacuum gap (layer 2), a receiving VO$_2$ film (layer 3), and the receiver substrate (layer 4). The film thickness of the emitting and receiving VO$_2$ is respectively $t_1$ and $t_3$, while the vacuum gap distance is represented as $d$. The emitter and receiver are assumed at respective local thermal equilibrium, whereas the latter is kept at a temperature of $T_L = 300$ K, at which VO$_2$ is always an insulator. On the other hand, the emitter temperature $T_H$ will be varied to be below (i.e., the "on" mode of the thermal switch) or above (i.e., the "off" mode) the VO$_2$ phase transition temperature at 341 K. A switching factor can be defined as:

$$\Phi = 1 - q''_{\text{off}} / q''_{\text{on}} \tag{2}$$

where $q''_{\text{on}}$ and $q''_{\text{off}}$ refer to the net heat flux for the "on" and "off" modes of thermal switch, respectively. Fluctuational electrodynamics [25], based on the stochastic nature of thermal emission, was used to calculate the near-field radiative heat fluxes for both "on" and "off" modes of the proposed vacuum thermal switch. The analytical expression for the radiative heat flux between two thin films with substrates at temperatures of $T_H$ and $T_L$ is [9, 26]

$$q'' = \frac{1}{\pi^2} \int_0^\infty d\omega \left[ \Theta(\omega, T_H) - \Theta(\omega, T_L) \right] \int_0^\infty s(\omega, \beta) d\beta \qquad (3)$$

Here, $\Theta(\omega, T) = \hbar\omega / \left[ \exp(\hbar\omega / k_B T) - 1 \right]$ is the mean energy of a Planck oscillator at thermal equilibrium temperature $T$, where $\hbar$ and $k_B$ are the reduced Planck constant and the Boltzmann constant, respectively. The exchange function $s(\omega, \beta)$ for propagating waves ($\beta < \omega/c$) and evanescent waves ($\beta > \omega/c$) is different and can be written as

$$s_{prop}(\omega, \beta) = \frac{\beta \left(1 - \left| R_{210}^s \right|^2 \right) \left(1 - \left| R_{234}^s \right|^2 \right)}{4 \left| 1 - R_{210}^s R_{234}^s e^{i 2\gamma_2 d} \right|^2} + \frac{\beta \left(1 - \left| R_{210}^p \right|^2 \right) \left(1 - \left| R_{234}^p \right|^2 \right)}{4 \left| 1 - R_{210}^p R_{234}^p e^{i 2\gamma_2 d} \right|^2} \qquad (4a)$$

and

$$s_{evan}(\omega, \beta) = \frac{\text{Im}(R_{210}^s) \text{Im}(R_{234}^s) \beta e^{-2\text{Im}(\gamma_2)d}}{\left| 1 - R_{210}^s R_{234}^s e^{i 2\gamma_2 d} \right|^2} + \frac{\text{Im}(R_{210}^p) \text{Im}(R_{234}^p) \beta e^{-2\text{Im}(\gamma_2)d}}{\left| 1 - R_{210}^p R_{234}^p e^{i 2\gamma_2 d} \right|^2} \qquad (4b)$$

where $\beta$ is the parallel-component wavevector, $\gamma$ is the vertical-component wavevector, $R_{210(234)}$ is respectively the reflection coefficient from the vacuum gap to either emitter (i.e., $R_{210}$) or to the receiver (i.e., $R_{234}$), and can be calculated by thin-film optics for both s and p polarizations [27, 28]

$$R_{234(210)} = \frac{r_{23(21)} + r_{34(10)} e^{i 2\gamma_{3(1)} t_{3(1)}}}{1 + r_{23(21)} r_{34(10)} e^{i 2\gamma_{3(1)} t_{3(1)}}} \qquad (5)$$

where $r_{ij}$ is the Fresnel reflection coefficient at the single interface from medium $i$ to medium $j$. while the subscripts 0, 1, 2, 3, 4 represent the emitter substrate, VO$_2$ emitter, vacuum gap, VO$_2$ receiver and receiver substrate, respectively, as indicated in Fig. 2. Note that when the thin films (i.e., layer 1 or 3) become semi-infinite, the thin-film reflection coefficient $R_{210}$ (or $R_{234}$) will be replaced by the reflection coefficient $r_{21}$ ($r_{23}$) at a single interface. In addition, if the film substrate (i.e., layer 0 or 4) is vacuum, the energy associated with propagating waves will not be absorbed when it transmits into the vacuum substrate. In this case, $(1-R_{ij})$ will be replaced by $(1-R_{ij}-T_{ij})$ in Eq. (4a), where $T_{ij}$ is the transmission coefficient from medium $i$ to thin-film layer $j$, obtained by thin-film optics [27, 28]

$$T_{234(210)} = \frac{t_{23(21)}t_{34(10)}e^{i\gamma_{3(1)}t_{3(1)}}}{1 + r_{23(21)}r_{34(10)}e^{i2\gamma_{3(1)}t_{3(1)}}} \qquad (6)$$

where $t_{ij}$ is the transmission coefficient at the single interface from medium $i$ to medium $j$.

When VO$_2$ is metallic, which is an isotropic medium, the reflection coefficient at the VO$_2$-vacuum interface can be calculated by isotropic wave propagation [29]. However, when VO$_2$ is in the insulating phase, it is a uniaxial insulator, and anisotropic wave propagation should be considered. Taking the same consideration with the structures of aligned nanowire array [30, 31] and nanoporous SiC plate [32], the ordinary component $\varepsilon_O$ and extraordinary component $\varepsilon_E$ of dielectric function are used to calculate the reflection coefficient [31]. For s polarized wave, the electric field is always perpendicular to the optical axis, and thus only ordinary response is considered such that $r_{ij}^s = (\gamma_i^s - \gamma_j^s)/(\gamma_i^s + \gamma_j^s)$, $t_{ij}^s = 2\gamma_i^s/(\gamma_i^s + \gamma_j^s)$, where $\gamma_i^s = \sqrt{\varepsilon_i^O(\omega/c)^2 - \beta^2}$ and $i$ refers to layer index [29, 32, 33], where the superscript "s" refers to the s polarization. While for p polarized wave, both ordinary and extraordinary dielectric responses have to be considered since the electrical field would always have components perpendicular to and parallel to the optics axis of VO$_2$ insulator. The reflection coefficient for p polarized waves can be obtained from $r_{ij}^p = (\varepsilon_j^O\gamma_i^p - \varepsilon_i^O\gamma_j^p)/(\varepsilon_j^O\gamma_i^p + \varepsilon_i^O\gamma_j^p)$, $t_{ij}^p = 2\varepsilon_j^O\gamma_i^p/(\varepsilon_j^O\gamma_i^p + \varepsilon_i^O\gamma_j^p)$, where $\gamma_i^p = \sqrt{\varepsilon_i^O(\omega/c)^2 - \varepsilon_i^O\beta^2/\varepsilon_i^E}$ [32, 33]. Note that the expressions are also valid for an isotropic medium, as $\varepsilon_O$ and $\varepsilon_E$ are identical.

## 3. Results and Discussion

### 3.1 Near-field Radiative Heat Fluxes with Different Emitter Temperatures

Let us first consider the 3-layer configuration of the proposal near-field thermal switch made of semi-infinite $VO_2$ emitter and receiver, as shown in the inset of Fig. 3, which presents the net radiative heat flux between the $VO_2$ plates separated by a vacuum gap of $d$ = 50 nm, when the emitter temperature $T_H$ changes from 330 K to 350 K. The $VO_2$ emitter is insulating at the beginning when the temperature is below 341 K, while the $VO_2$ receiver, whose temperature is fixed at $T_L$ = 300 K, is always an insulator. The radiative heat flux is 15 kW/m$^2$ at $T_H$ = 330 K, increases almost linearly with the emitter temperature, and reaches 21 kW/m$^2$ at 341 K. However, when the emitter temperature exceeds 341 K, the emitter becomes metallic. As a result, the radiative heat flux decreases abruptly to 5.2 kW/m$^2$ at $T_H$ = 342 K, and increases slightly to 6.3 kW/m$^2$ at 350 K. Therefore, it can be clearly seen that, when crossing the phase transition of $VO_2$ at 341 K, near-field radiative heat transfer is significantly reduced from 21 kW/m$^2$ reduces to 5 kW/m$^2$, resulting in a switching factor of 76.2% at $d$ = 50 nm. When the emitter temperature is below the phase transition point, it can be treated as the "on" mode for this vacuum thermal switch made of $VO_2$, while the "off" mode can be activated when the temperature is above the phase transition.

In order to have a better understanding in the physical mechanism of thermal switching effect realized by the near-field radiative transfer between two $VO_2$ plates, the contour of exchange function $s(\omega,\beta)$ is plotted for both "on" and "off" modes as shown in Fig. 4 at the same vacuum gap distance ($d$ = 50 nm). The color bar on the right shows the scale for $s(\omega,\beta)$ with the brightest color representing the peak value. As shown in Fig. 4(a) for the "on" mode, strong enhancement suggested by the bright contour can be clearly seen, with a major one around $\omega_m =$ $1.5 \times 10^{14}$ rad/s and several other peaks at smaller angular frequencies. The enhancement in the $s$ function, which results in the enhancement of radiative heat flux, is due to the strong phonon-phonon coupling, i.e., the excitation of SPhPs, between the insulating VO$_2$ plates across the small vacuum gap. Note that, for the "on" mode, both the emitter and the receiver are insulating VO$_2$ with several phonon modes in the infrared region. Those phonons could constructively resonant across the vacuum gap, resulting in increases of radiative heat transfer modes in the near field. The strongest enhancement occurs at $\omega_m = 1.5 \times 10^{14}$ rad/s in the $s$ function contour. On the other hand, when the emitter becomes metallic, which is the case for the "off" mode, the $s$ function between an insulating VO$_2$ and a metallic VO$_2$ is greatly suppressed as shown in Fig. 4(b), in comparison to Fig. 4(a). The phonon modes of the insulating VO$_2$ could not constructively couple to the other metallic interface, which does not support any phonon modes. Therefore, the radiative heat transfer is greatly reduced for the "off" mode, resulting in strong thermal switching effect with such a vacuum thermal switch made of VO$_2$.

### 3.2 Thermal Switching Factors at Different Vacuum Gap Distances

To have a better idea on how the thermal switching effect varies with vacuum gap distances, Figure 5 plots the total net heat fluxes with different vacuum gap distances for both "on" and "off" modes, as well as the switching factor defined in Eq. (2). Here, the "on" and "off" modes respectively correspond to the emitter temperature $T_H$ of 341 K and 342 K, while the receiver is fixed at $T_L = 300$ K. The switching factor can be as large as 0.85 when gap distance is smaller than 10 nm, indicating that 85% of heat flux could be cut off when the thermal switch turns from "on" mode to "off" mode. In addition, the large heat flux between two VO$_2$ plates is more than 500 kW/m$^2$ for the "on" mode at sub-10-nm vacuum gaps, suggesting potential applications in microelectronic cooling. Since the dependence of the near-field radiative transfer on the vacuum gaps is different between "on" and "off" modes, the switching factor starts to decrease when the gap distance is larger than 30 nm, reaches a minimum of 0.46 around $d = 300$ nm, and increases again to 0.51 at vacuum gaps of 1 μm.

### 3.3 Penetration Depth inside Uniaxial Insulating VO$_2$

To obtain an insight on how the near-field radiative energy is transferred, the z component of Poynting vector $S_z(\omega,z)$ is plotted in Fig. 6 at different depths inside the insulating VO$_2$ emitter and receiver separated by a vacuum gap of 50 nm. The penetration depth inside isotropic media has already been investigated in Refs. [34, 35], but when it comes to a uniaxial medium, we need to separate s and p polarizations and apply different expressions of vertical wavevector and reflection coefficient as described above. The Poynting vector $S_z(\omega,z)$ is normalized to that in the vacuum, and is calculated at several angular frequencies of $0.7 \times 10^{14}$ rad/s, $1.05 \times 10^{14}$ rad/s and $1.5 \times 10^{14}$ rad/s, at which coupled SPPs occur as shown in Fig. 4(a). From Fig. 6, we can clearly see that the penetration depth, i.e., a distance at which the radiation power is attenuated by a factor of 1/e (approximately 37%), varies with different resonance frequencies. The penetration depth for the spectral energy at $\omega = 0.7 \times 10^{14}$ rad/s is the largest, which is about 10 times of vacuum gap distance, while that for resonance frequency of $1.05 \times 10^{14}$ rad/s is the shortest, which is about the same with vacuum gap distance. The penetration depth for total radiative energy after the integration over $\omega$ is about 3 times of vacuum gap distance, similar to that for spectral energy of resonance frequency at $1.5 \times 10^{14}$ rad/s. This is because most of energy transfer is concentrated at this resonance frequency. With a thickness of several times of the penetration depth, an insulating $VO_2$ film can be reasonably treated semi-infinite. On the other hand, if the film thickness is comparable or smaller than the penetration depth, the thin-film wave propagation has to be considered, which could influence the near-field radiative transfer and thereby possibly further enhance the thermal switching effect.

### 3.4 Thermal Switching Effect with a Thin-film Receiver

To investigate the thin-film effect on the near-field radiative thermal switching, a 4-layer configuration with a thin-film receiver with thickness $t_3$, as shown in the inset of Fig. 7, is considered. The near-field heat transfer along with the switching factor between a semi-infinite $VO_2$ emitter and a free-standing $VO_2$ thin-film receiver (i.e., vacuum substrate) is calculated as a function of the film thickness for both the "on" (i.e., $T_H = 341$ K) and "off" modes (i.e., $T_H = 342$ K). The gap distance is considered to be $d = 300$ nm, at which the switching factor has a minimum between two semi-infinite $VO_2$ plates as shown in Fig. 5, aiming to further enhance thermal switching effect with the thin-film effect.

As seen in Fig. 7, both the radiative heat fluxes for the "on" and "off" modes increase monotonically when the film thickness increases from 1 nm to 10 μm, which is simply due to the increase in the volume of receiving material. At the vacuum gap of 300 nm, a uniaxial insulating $VO_2$ film with 10-μm thickness can still not be considered to be semi-infinite. For thin-film thickness smaller than the penetration depth, the SPhPs at both interfaces of thin-film $VO_2$ receiver can also couple with each other inside the thin film, in addition to the coupled SPhPs across the vacuum gap between two insulating $VO_2$. When $t_3$ increases from 1 nm to 20 nm, both the heat fluxes have the same dependence on the film thickness, which results in a nearly constant switching factor of $\Phi=0.7$. For sub-20-nm films, when the thickness is much smaller than the vacuum gap distance (i.e. $t_3 \ll d$ ), the SPhPs coupling inside the thin film will dominate near-field heat transfer [36], for which the splitting of the near-field heat flux into symmetric and anti-symmetric resonant modes could increase the number of modes in the near field. When $t_3 >$ 20 nm, the coupling between symmetric and anti-symmetric modes becomes weaker and the coupling of evanescent waves inside the vacuum gap becomes stronger as the film thickness increases. Thus, the radiative heat flux at the "on" mode starts to increase with a smaller slope around $t_3 = 50$ nm, while this occurs for the "off" mode around $t_3 = 1$ μm or so. As a result, the switch factor $\Phi$ starts to experience a reduction with a VO$_2$ thin film at $t_3 = 50$ nm and reaches a minimum of 0.39 at $t_3 = 1$ μm. When the film thickness further increases, the thin-film effect is vanishing and the heat transfer approaches to that between two bulks. The switching factor $\Phi$ increases to 0.44 for a 10-μm-thick VO$_2$ film. Clearly, the switching factor $\Phi$ becomes larger when the receiving VO$_2$ made of a thin film with thickness below 50 nm than a semi-infinite plate.

The dependence of the film thickness clearly suggests further enhancement of the switching effect by using a thin film receiver. Figure 8 compares the switching factors from a vacuum thermal switch in a 3-layer configuration with those from a 4-layer configuration with a thin-film receiver of 20 nm, at a wide range of vacuum gap distances. Clearly, when using a thin-film receiver, the thermal switching effect is greatly enhanced over a broad range of vacuum distances from 10 nm to nearly 1 μm. A maximal switching factor over 0.83 can be achieved at sub-100-nm gap. Moreover, the switching factor $\Phi$ is improved from 0.62 to 0.83 at $d = 100$ nm, and more than 50% increase is obtained at $d = 200$ nm, where the switching factor is improved from 0.49 to 0.78. Note that, vacuum gaps of a few hundred of nanometers could be practically achievable and the optimized vacuum thermal switch with a thin-film receiver would facilitate practical applications in thermal management and thermal circuits.

*3.5 Effect of SiO₂ substrate on near-field thermal switching*

From a practical perspective, the thin-film VO₂ receiver has to be deposited on a real substrate rather thermal vacuum. Therefore, the effect of film substrate needs to be studied for a practical design of a vacuum thermal switch made of VO₂. Here, we consider SiO₂ as the substrate material for the VO₂ films as an example to illustrate the purpose here, while other substrate materials like Al₂O₃, silicon, and even metals can be analyzed in a similar approach. The optical properties of SiO₂ were taken Palik's data [37]. Besides considering a 4-layer vacuum thermal switch with a thin receiving film on SiO₂ substrate, a 5-layer vacuum thermal switch design with both thin-film emitter and receiver on SiO₂ substrate will be also investigated. The theoretical model described in Section 2 will be employed for calculating the near-field radiative heat transfer between two uniaxial films on substrates.

With SiO₂ substrate, the receiving film thickness ($t_3$) effect on the radiative heat fluxes for both "on" and "off" modes as well as the switching factor at the vacuum gap of $d = 300$ nm is shown in Fig. 9(a). Similar to the case with vacuum substrate shown in Fig. 7, the heat fluxes for both "on" and "off" modes will increase with thicker films. However, due to the presence of SiO₂ substrate, the growth rate (i.e., the slope) of heat flux as a function of $t_3$ becomes smaller, while the amplitude of heat fluxes is higher, which can be understood by the modified reflection coefficients across the thin receiving film due to the SiO₂ substrate. Moreover, SiO₂ is a polar material with strong phonon modes in the infrared region, which also helps the energy absorption in addition to the VO₂ thin film. As a result, the thermal switching factor Φ increases slightly to

0.48 with the $SiO_2$ substrate at $d$ = 300 nm from 0.45 with vacuum substrate at the same receiving film thickness $t_3$ = 50 nm.

Now with the receiving film thickness $t_3$ = 50 nm, the thin-film and substrate effects for the emitter at the same vacuum gap distance are also investigated. Figure 9(b) plots the radiative heat fluxes at varying emitting film thickness $t_1$ for the proposed 5-layer radiative thermal switch design as illustrated in the inset. The radiative heat transfer is enhanced with a thin-film emitting $VO_2$ on SiO2 substrate, in particular, for the "off" mode for which the emitting $VO_2$ film is metallic. More interestingly, the radiative heat flux for the "off mode" with $T_H$ = 342 K actually becomes larger than that for the "on" mode with $T_H$ = 341 K with ultrathin emitting films when $t_1$ < 10 nm. Note that, as both the emitter and receiver are on $SiO_2$ substrate, the coupling between the emitting $SiO_2$ substrate to the receiver dominates the near-field radiative transport across the vacuum gap when the emitting $VO_2$ is extremely thin. Therefore, the optical property change due to the phase transition of the ultrathin $VO_2$ emitting film has negligible effect on the near-field heat transfer. The heat flux becomes larger in the "off" mode is simply due to the higher emitter temperature, resulting in negative switching factors. However, when the emitting $VO_2$ film is thicker, the coupling of the VO2 film to the receiver becomes dominating. As a result, radiative heat fluxes in both modes decreases, and more importantly, the heat flux at the "off" mode decreases in a faster pace, which can be easily understood by the growing phase transition effect of VO2 on modulating radiative heat transfer. The switching factor Φ increases with larger emitting film thickness, reaches a maximum of almost 0.5 at $t_1$ = 500 nm, and then saturates around 0.48, indicating the effect of film thickness disappears and approaching the semi-infinite emitting $VO_2$ case. The effects of thin-film and substrates provide practical insights for experimental demonstration and further optimization of proposed near-field radiative thermal switch in the future.

## 4. Conclusions

In summary, strong thermal switching effect using the phase transition material $VO_2$ is theoretically demonstrated based on near-field radiative heat transfer between phase transition material $VO_2$ separated by nanometer vacuum gaps. The radiative heat flux is significantly reduced when the $VO_2$ emitter changes from insulator to metal across the phase transition temperature of 341 K, switching the heat flow from the "on" to "off" mode. The switching factor between two semi-infinite $VO_2$ plates is as large as 0.85 when the gap distance is smaller than 10 nm, but is as small as 0.46 at $d = 300$ nm. The penetration depth of near-field radiation in the uniaxial $VO_2$ insulators indicates the effect of $VO_2$ thickness on the radiative heat fluxes, and the thermal switching factor is further enhanced with a thin receiver of several tens of nanometers thick in a broad range of vacuum gaps. The $SiO_2$ substrate effect has also been investigated for a practical 4-layer configuration as well as a 5-layer design with both thin-film $VO_2$ emitter and reciever. The results and insights gained from this work are of practical importance for thermal management and even achieving thermal circuit in the future, thus provide guideline for future design and experimental demonstration of efficient vacuum thermal switches.

**Acknowledgements**:

YY and LW thank the support from the ASU New Faculty Startup program.

**Figure Captions:**

Fig. 1 The dielectric functions of VO$_2$: (a) real part ($\varepsilon'$) and (b) imaginary part ($\varepsilon''$). The insulating VO$_2$ is a uniaxial medium with ordinary ($\varepsilon_\mathrm{O}$) and extraordinary dielectric functions ($\varepsilon_\mathrm{E}$), while the metallic VO$_2$ is an isotropic medium with dielectric function $\varepsilon_\mathrm{m}$. The optical constants of VO$_2$ are taken from Ref. [38].

Fig. 2 Schematic of a vacuum thermal switch made of five layers with two VO$_2$ thin films on substrates separated by a vacuum gap $d$. The emitter and receiver VO$_2$ film thickness are denoted as $t_1$ and $t_3$, respectively. The emitter temperature is $T_\mathrm{H}$, which varies above 300 K, while the receiver is $T_\mathrm{L}$ kept constant at 300 K. The directions of wavevectors are also shown in this figure.

Fig. 3 The total radiative heat fluxes between two semi-infinite VO2 plates, as shown in the inset figure, with different emitter temperatures at the gap distance of 50 nm with the receiver temperature fixed at 300 K. The thermal switch is in "on" mode when emitter temperature is below 341 K, while it turns to "off" mode when emitter temperature is above 341 K.

Fig. 4 Contour plots of the exchange function $s(\omega,\beta)$ between two semiinfinite VO$_2$ plates separated by a vacuum gap of $d$ = 50 nm for (a) "on" mode with two insulating VO$_2$ layers and (b) "off" mode with one insulating VO$_2$ layer and one metallic VO$_2$ layer.

Fig. 5 The total heat fluxes in both "on" and "off" modes and the resulting switching factor at different vacuum gap distances. The receiver temperature is 300 K. The emitter temperature is set as 341 K and 342 K for the "on" and "off" modes, respectively.

Fig. 6 The field distribution of the spectral Poynting vector (z component) near the surfaces of the emitter and receiver at several phonon frequencies of insulating VO$_2$ as well as the total one.

The Poynting vector is normalized to that in the vacuum and the two semi-infinite insulating VO$_2$ plates is separated by $d$ = 50 nm.

Fig. 7  The film thickness effect of the VO$_2$ receiver on the radiative heat fluxes for both "on" and "off" modes as well as the switching factor at the vacuum gap of 300 nm. Inset is the schematic of the 4-layer vacuum thermal switch consisting of a VO$_2$ thin film receiver with thickness $t_3$ and a semi-infinite VO$_2$ emitter.

Fig. 8  Comparison on the switching factor between the 3-layer vacuum switch and the 4-layer one with a 20-nm-thick free-standing thin-film receiver at different vacuum gaps. The temperature of the emitter is set as 341 K and 342 K for the "on" and "off" modes, respectively.

Fig. 9 (a) The film thickness effect of the VO$_2$ receiver on the radiative heat fluxes for both "on" and "off" modes as well as the switching factor at the vacuum gap of 300 nm with SiO$_2$ substrate. (b)  The film thickness effect of the VO$_2$ emitter on the radiative heat fluxes for both "on" and "off" modes as well as the switching factor at the vacuum gap of 300 nm and with receiver thickness of 50 nm, for which SiO$_2$ substrates are applied for both emitter and receiver.

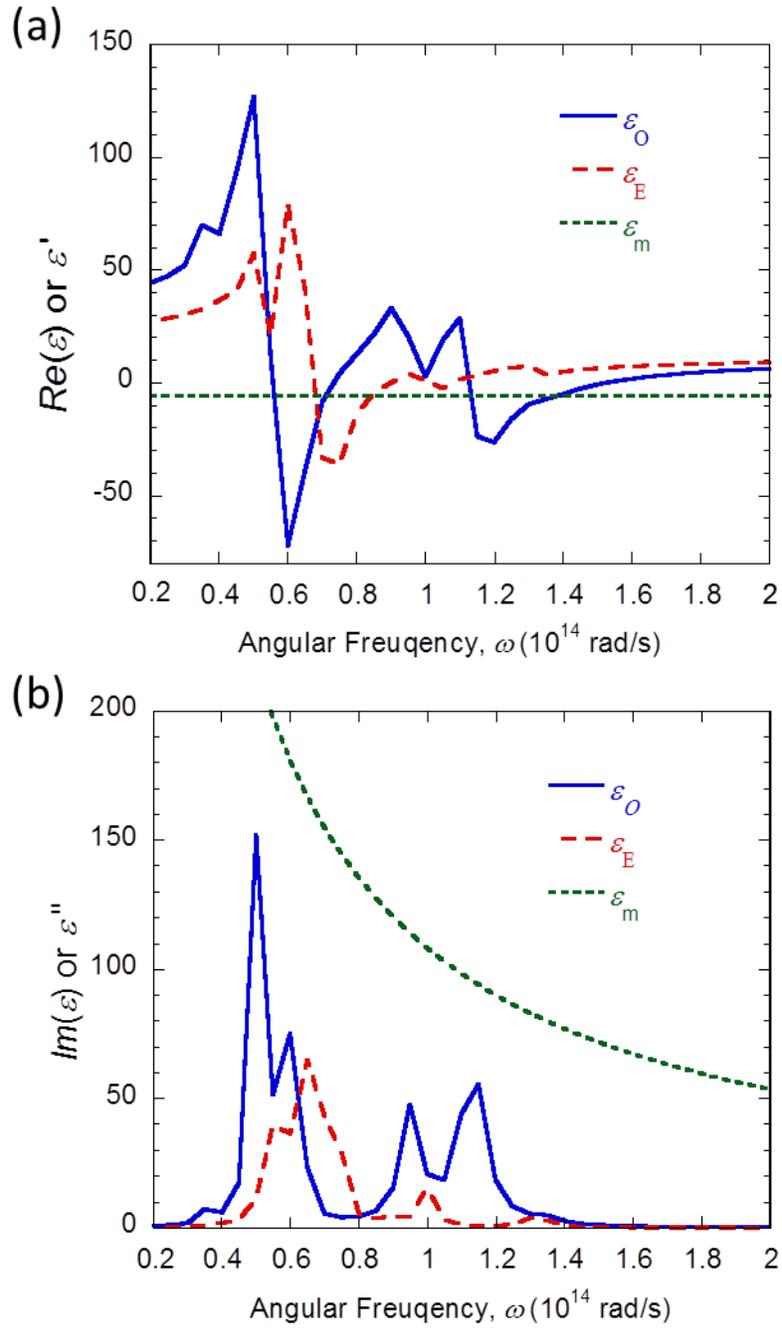

Yang et al., Fig. 1

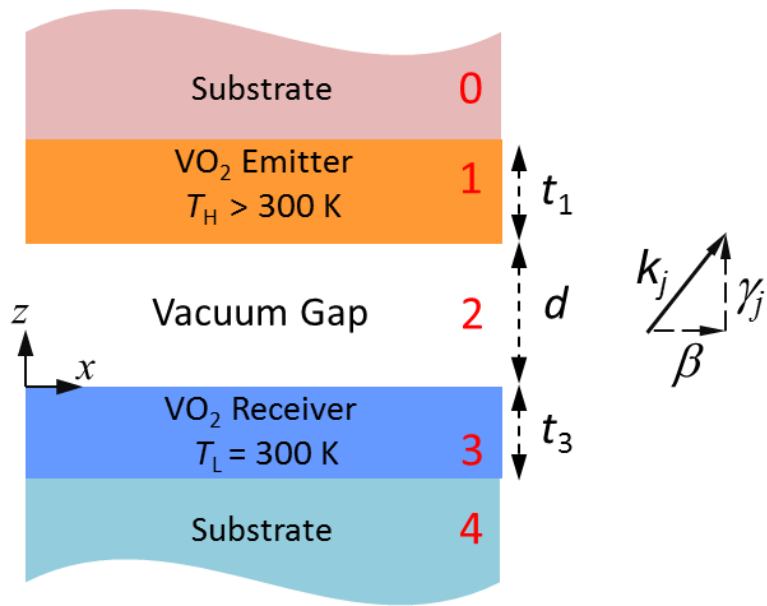

Yang et al., Fig. 2

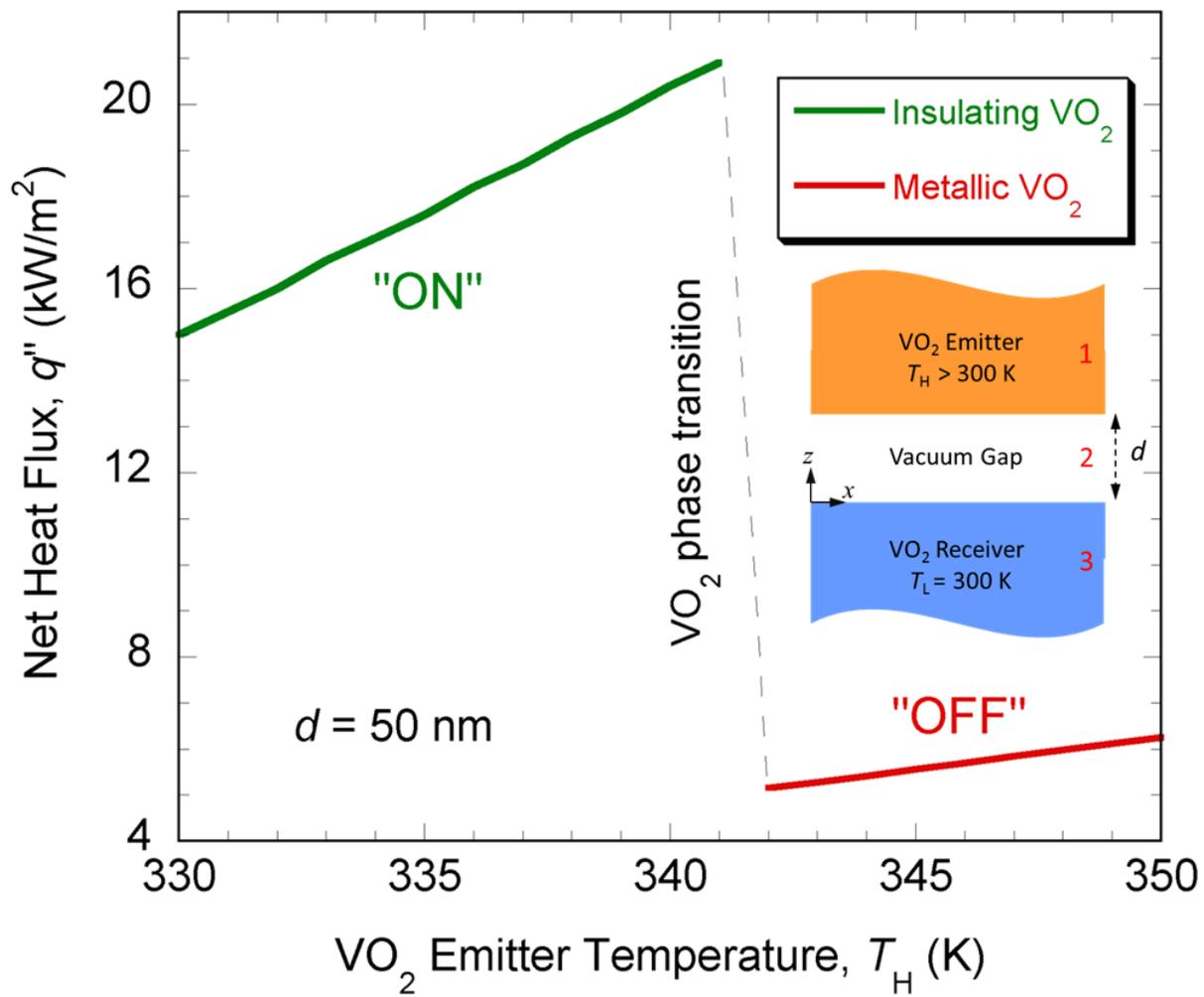

Net Heat Flux, $q''$ (kW/m$^2$) vs. VO$_2$ Emitter Temperature, $T_H$ (K)

Insulating VO$_2$
Metallic VO$_2$

"ON"

VO$_2$ phase transition

$d$ = 50 nm

"OFF"

VO$_2$ Emitter $T_H > 300$ K  1
Vacuum Gap  2  $d$
VO$_2$ Receiver $T_L = 300$ K  3

Yang et al., Fig. 3

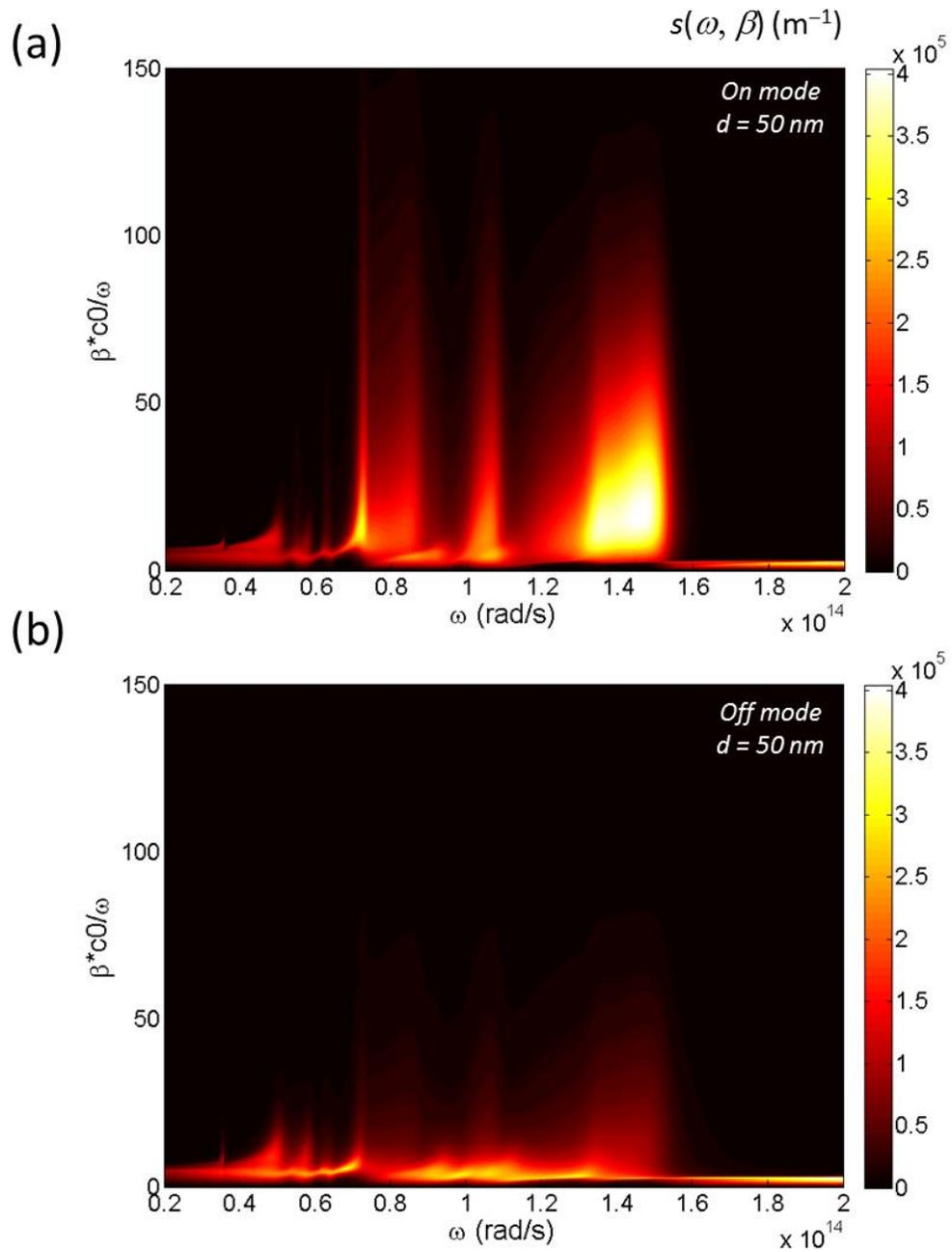

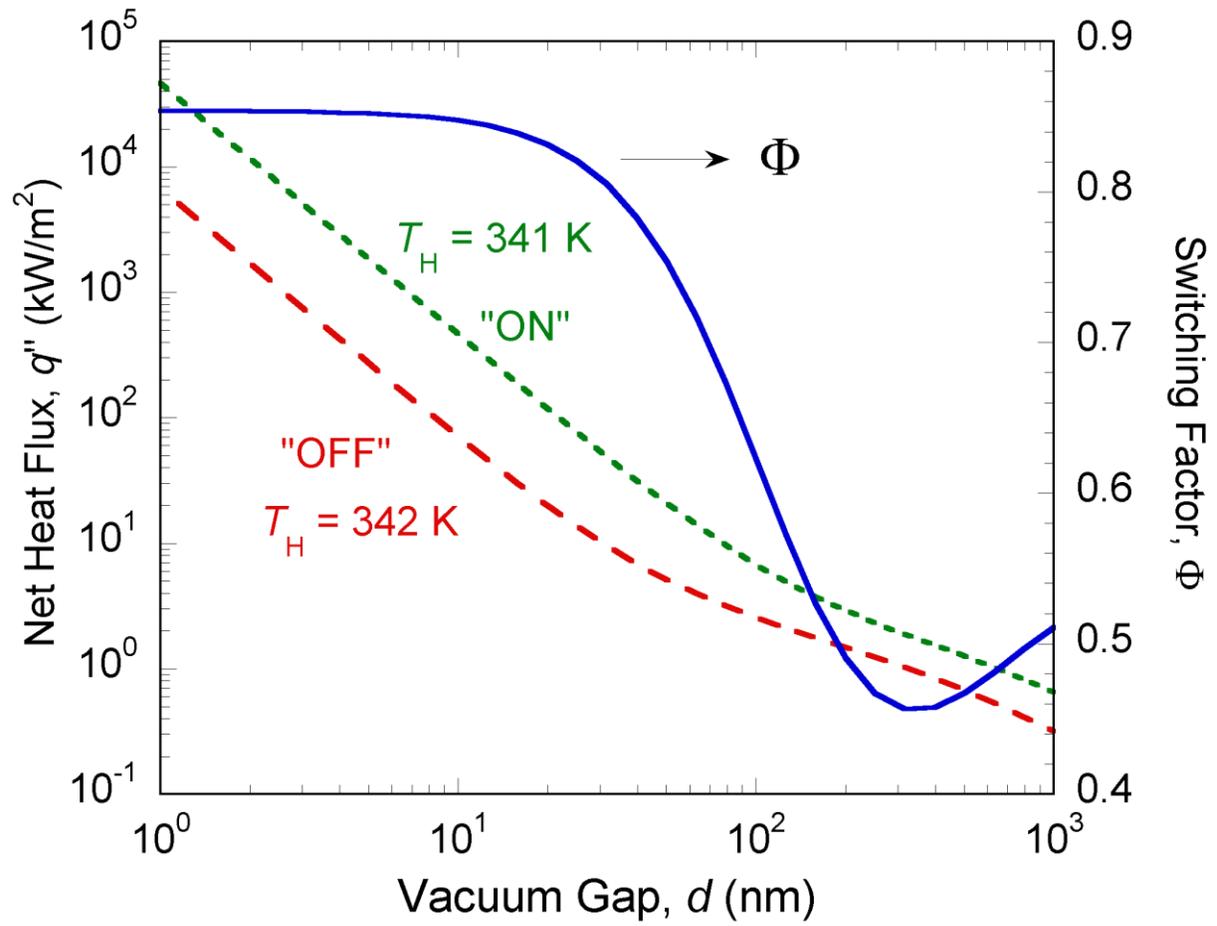

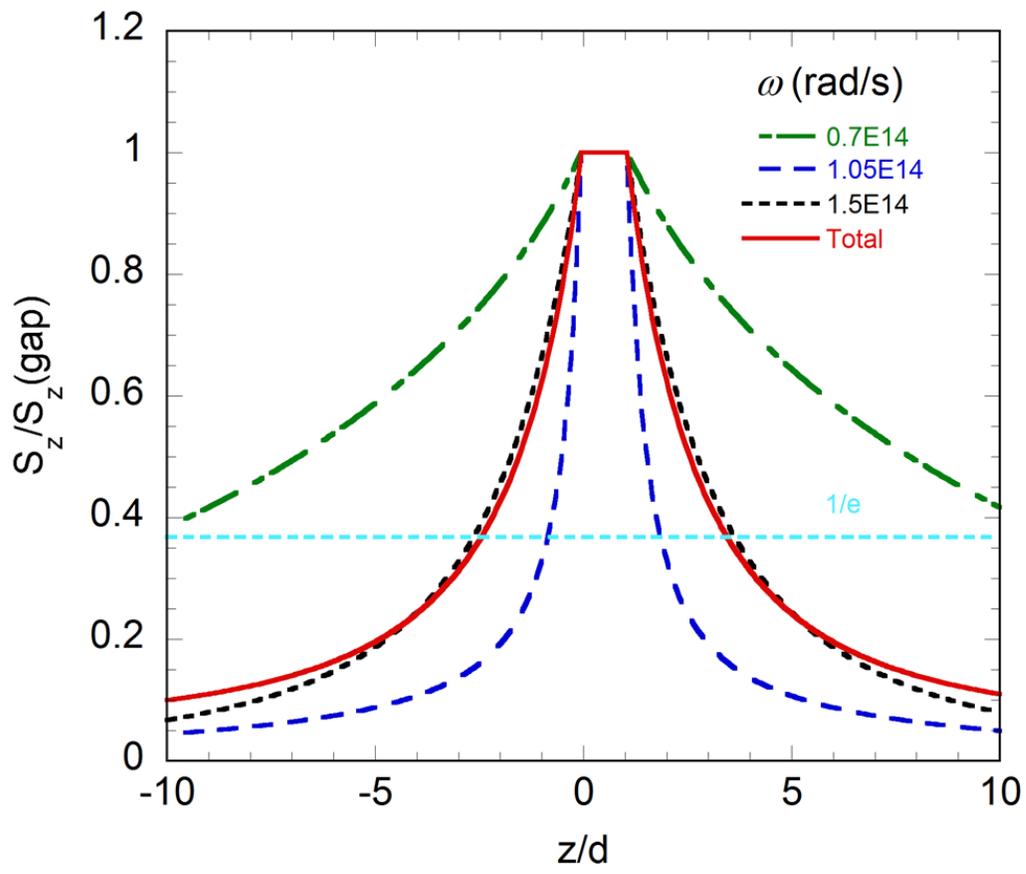

Yang et al., Fig. 6

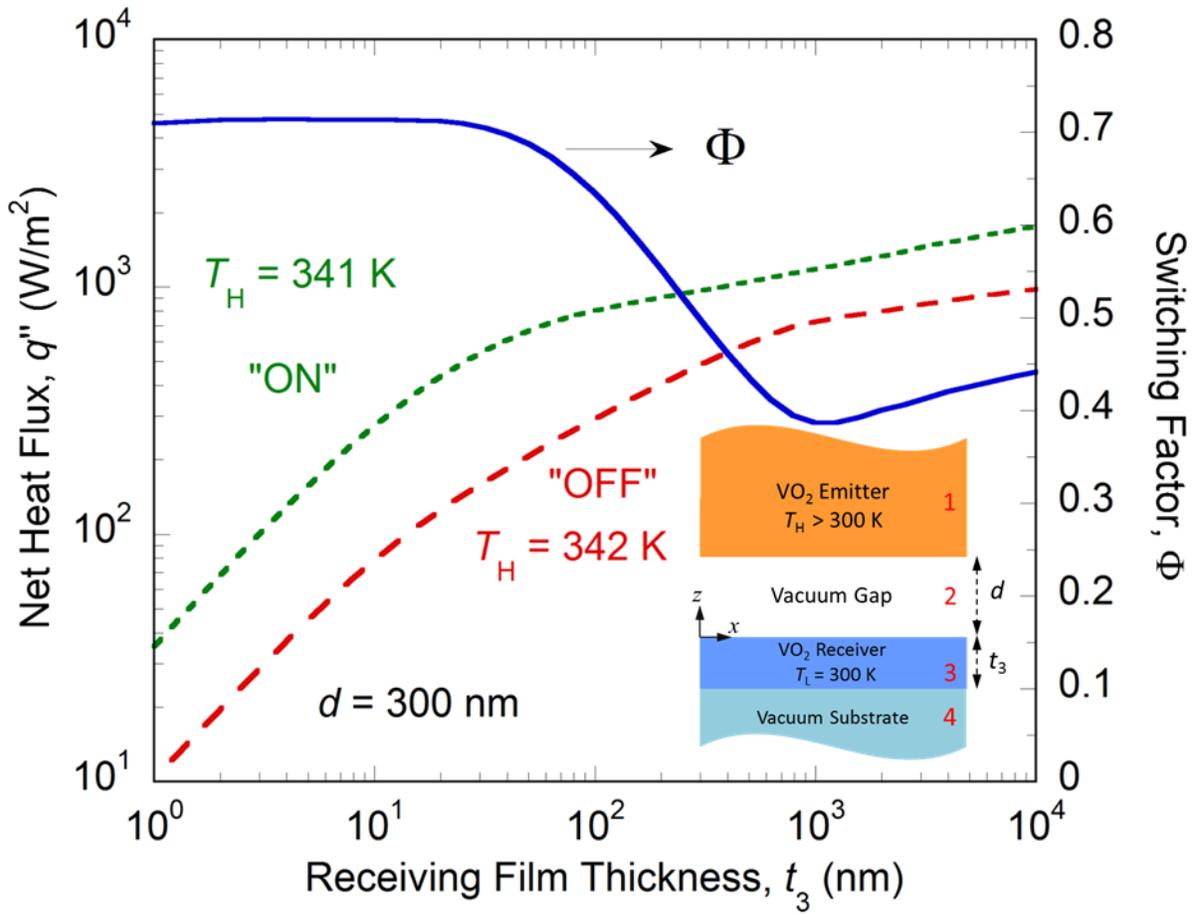

Yang et al., Fig. 7

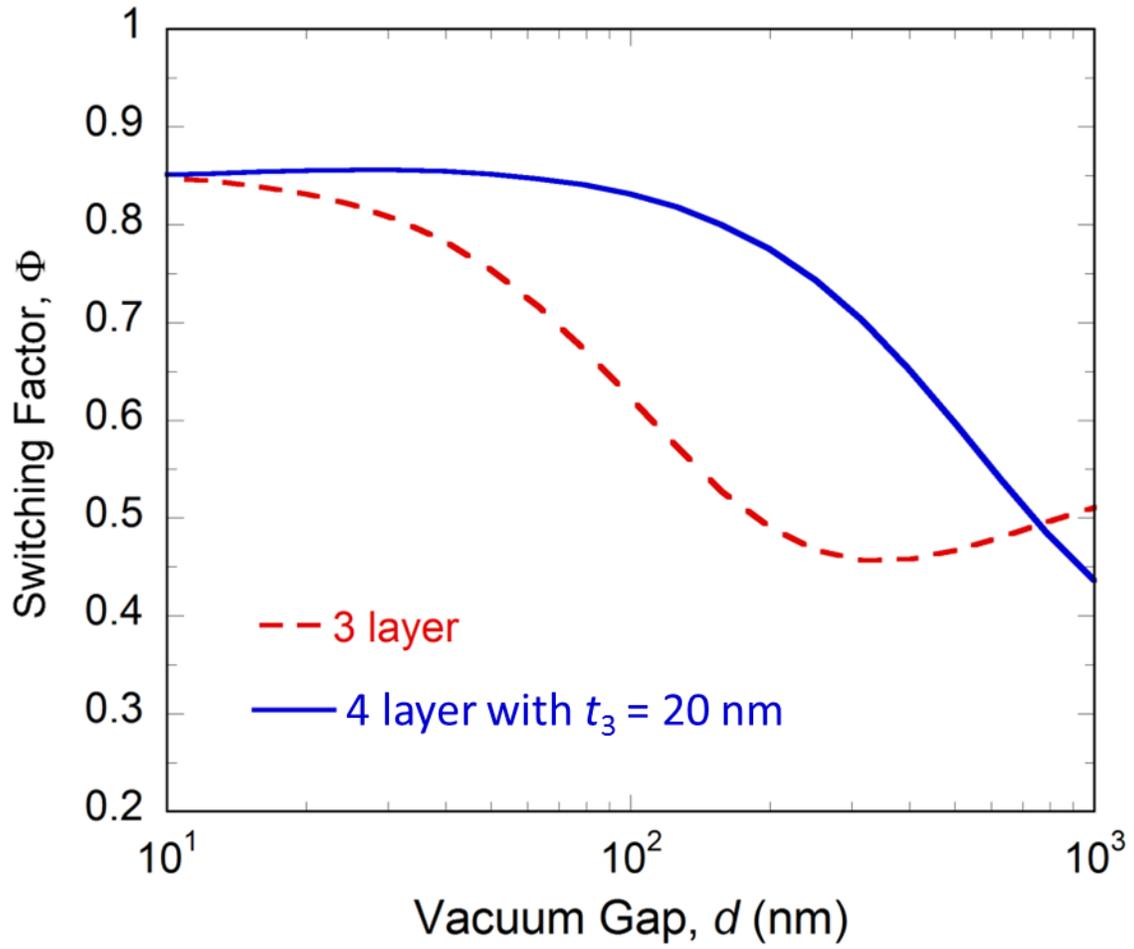

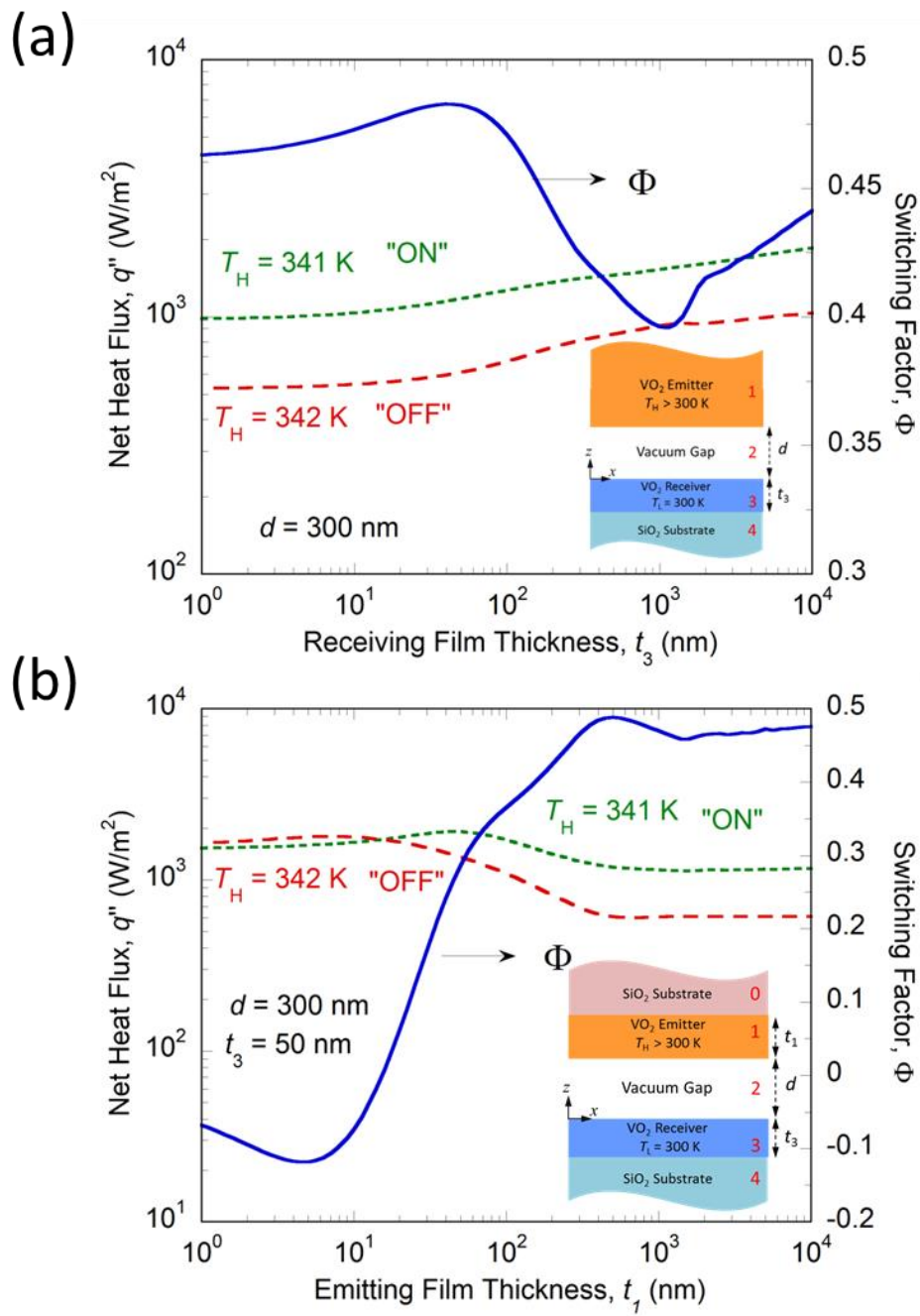

Yang et al., Fig. 9